\documentclass[aps,prb,twocolumn,showpacs,citeautoscript,reprint,longbibliography,superscriptaddress]{revtex4-1}
\usepackage{graphicx}
\usepackage{bm}
\usepackage{natbib}
\usepackage{amsmath}
\usepackage{xfrac}
\usepackage{nicefrac}
\usepackage[colorlinks=true, urlcolor=blue, linkcolor=blue, citecolor=blue, pdfborder={0 0 0}]{hyperref}
\usepackage[caption=false]{subfig}
\usepackage{cleveref}
\usepackage{dcolumn}
\usepackage{ulem}

\crefname{figure}{Fig.}{Figs}
\crefname{table}{Table}{Tables}

\begin{document}

\title{Disorder Dependence of Interface Spin Memory Loss}
	
\author{Kriti Gupta}
\affiliation{Faculty of Science and Technology and MESA$^+$ Institute for Nanotechnology, University of Twente, P.O. Box 217,
		7500 AE Enschede, The Netherlands}
\author{Rien J.H. Wesselink}
\affiliation{Faculty of Science and Technology and MESA$^+$ Institute for Nanotechnology, University of Twente, P.O. Box 217, 7500 AE Enschede, The Netherlands}
\author{Ruixi Liu}
\affiliation{The Center for Advanced Quantum Studies and Department of Physics, Beijing Normal University, 100875 Beijing, China}
\author{Zhe Yuan}
\email[Email: ]{zyuan@bnu.edu.cn}
\affiliation{The Center for Advanced Quantum Studies and Department of Physics, Beijing Normal University, 100875 Beijing, China}
\author{Paul J. Kelly\thanks{corresponding author}}
\email[Email: ]{P.J.Kelly@utwente.nl}
\affiliation{Faculty of Science and Technology and MESA$^+$ Institute for Nanotechnology, University of Twente, P.O. Box 217,
	7500 AE Enschede, The Netherlands}
\affiliation{The Center for Advanced Quantum Studies and Department of Physics, Beijing Normal University, 100875 Beijing, China}
	
\date{\today}
	
	
\begin{abstract}
The discontinuity of a spin-current through an interface caused by spin-orbit coupling is characterized by the spin memory loss (SML) parameter $\delta$. 
We use first-principles scattering theory and a recently developed local current scheme to study the SML for Au$|$Pt, Au$|$Pd, Py$|$Pt and Co$|$Pt interfaces. We find a minimal temperature dependence for nonmagnetic interfaces and a strong dependence for interfaces involving ferromagnets that we attribute to the spin disorder. The SML is larger for Co$|$Pt than for Py$|$Pt because the interface is more abrupt. Lattice mismatch and interface alloying strongly enhance the SML that is larger for a Au$|$Pt than for a Au$|$Pd interface. The effect of the proximity induced magnetization of Pt is negligible. 

\end{abstract}
	
\pacs{}
	
\maketitle

With the discovery of the giant magnetoresistance (GMR) effect in magnetic multilayers \cite{Baibich:prl88, Binasch:prb89}, it was recognized that interfaces play a key role in spin transport phenomena. In semiclassical formulations \cite{vanSon:prl87, Valet:prb93, Bass:jmmm16} of transport, these appear as discrete resistances and the description of the transport of electrons through a multilayer requires a resistivity $\rho$ for each material as well as a resistance $R_{\rm I}$ for each interface. Bulk resistivities are readily measured, interface resistances much less so. Magnetic materials require spin-dependent bulk resistivities $\rho_\sigma$ and interface resistances $R_\sigma$. Because spin is not conserved, describing its transport additionally requires a spin-flip diffusion length (SDL) $l_{\rm sf}$ in each material, as well as its counterpart  for each interface, the spin memory loss (SML) parameter $\delta$.  
Determining $l_{\rm sf}$ requires distinguishing interface and bulk contributions. Because doing so is non-trivial, the interface contribution had been largely neglected and values of $l_{\rm sf}$ reported over the last decade for well studied materials like Pt span an order of magnitude \cite{Bass:jpcm07, Sinova:rmp15, Wesselink:prb19}. 

Almost everything we know about interface parameters is from current-perpendicular-to-the-plane (CPP) magnetoresistance experiments \cite{Galinon:apl05, Bass:jpcm07, Bass:jmmm16} interpreted using the semiclassical Valet-Fert (VF) model \cite{Valet:prb93}. Though these experiments are relatively simple to interpret, they are restricted to low temperatures as they require superconducting leads \cite{Bass:jmmm16} while calculations have so far only addressed ballistic interfaces \cite{Schep:prb97, Stiles:prb00, Xia:prb01, Xu:prl06, Belashchenko:prl16, Dolui:prb17}. Because the vast majority of experimental studies in spintronics is carried out at room temperature, there is a need to know how transport parameters, in particular those describing interfaces, behave at finite temperatures. 

This need is accentuated by the huge interest in recent years \cite{Hoffmann:ieeem13, Sinova:rmp15} in the spin Hall effect (SHE) \cite{Dyakonov:pla71, Hirsch:prl99, Zhang:prl00} whereby a longitudinal charge current drives a transverse spin current in nonmagnetic materials, and in its inverse, the inverse SHE (ISHE). Determination of the spin Hall angle (SHA) $\Theta_{\rm sH}$ that measures the efficiency of the SHE is intimately connected with the SDL and, because an interface is always involved, with the SML \cite{Rojas-Sanchez:prl14}. When use is made of spin pumping and the ISHE \cite{Saitoh:apl06, Ando:prl08, Mosendz:prl10, Mosendz:prb10} or the SHE and spin-transfer torque (STT) \cite{Liu:prl11}, the interface in question is between ferromagnetic (FM) and nonmagnetic (NM) materials. When the nonlocal spin-injection method is used \cite{Kimura:prl07, Vila:prl07}, two interfaces are involved: an FM$|$NM interface to create a spin accumulation and an NM$|$NM$'$ interface to detect it. 
Progress has been made by recognizing that bulk parameters like $l_{\rm sf}$ and $\Theta_{\rm sH}$ are very sample dependent and that the SML plays a key role in determining their values \cite{Rojas-Sanchez:prl14, LiuY:prl14, Nguyen:prl16, Sagasta:prb16}.
Recent studies suggesting that measurements of the SHA may actually be dominated by interface effects \cite{Rojas-Sanchez:prl14, LiuY:prl14, WangL:prl16, Amin:prb16a} are stimulating attempts to tailor these \cite{Zhu:prb18, Zhu:prl19a, Berger:prb18b, Zhu:prb19, Zhu:prl19b}. 

This makes it crucial to have a way to independently determine interface parameters. We recently described a formalism to evaluate local charge and spin currents \cite{Wesselink:prb19} from the solutions of fully relativistic quantum mechanical scattering calculations \cite{Starikov:prb18} that include temperature-induced lattice and spin disorder \cite{LiuY:prb11, LiuY:prb15}. This yielded a layer-resolved description of spin currents propagating through atomic layers of thermally disordered Pt and Py that allowed us to unambiguously determine bulk transport properties \cite{footnote5}. By focussing on spin currents, we can straightforwardly evaluate all the parameters entering the Valet-Fert semiclassical formalism \cite{Valet:prb93} that is universally used to interpret experiment \cite{Bass:jmmm16}. 

In this paper we focus on interface transport properties and study realistic interfaces between thermally disordered materials. Typical 
structures used in  experimental studies of the SHE contain a heavy NM metal with strong SOC and a 3$d$ transition metal (TM) or TM alloy ferromagnet \cite{Rojas-Sanchez:prl14, Saitoh:apl06, Ando:prl08, Mosendz:prl10, Mosendz:prb10, Liu:prl11, ZhangW:natp15, Sinova:rmp15}. We will study (i) Au$|$Pt and Au$|$Pd interfaces to shed light on the role of SOC and roughness at interfaces involving heavy TMs and (ii) Py$|$Pt and Co$|$Pt interfaces to examine the role of the FM magnetization and disorder in determining interface parameters, as well as the temperature dependence of all these.

{\color{red}\it Method.---}We begin by solving the VF equations analytically for the spin accumulation $\mu_{si}(z)$ and spin current $j_{si}(z)$ in a metallic multilayer. The solution in each region $i$ involves two coefficients $A_i$ and $B_i$ that are determined by appropriate boundary conditions \cite{Valet:prb93}. For an NM$|$NM$'$ system we will require that $j_s(0)=1$ at the left-lead$|$NM interface corresponding to injecting a fully spin polarized current from the left lead and that $j_s(\infty)=0$ requiring the NM$'$ material to be much thicker than its $l_{\rm sf}$ value. The interface (I) is initally considered as a bulklike material with resistivity $\rho_{\rm I}$, SDL $l_{\rm I} \equiv l_{\rm sf}^{\rm I}$ and thickness $t$ so that at the NM$|$I and I$|$NM$'$ interfaces the spin accumulation and spin current are continuous \cite{Baxter:jap99, Eid:prb02}. We then eliminate the coefficients $A_{\rm I}$ and $B_{\rm I}$ and take the limit $t \rightarrow 0$ thereby defining the areal interface resistance $AR_{\rm I}$ and the SML parameter $\delta$ as
\begin{equation}
AR_{\rm I} = \lim_{t \to 0} \rho_{\rm I} t  \,\,\,\,\,\,{\rm and} \,\,\,\,\,\, \delta = \lim_{t \to 0} t/l_{\rm I} .
\label{eq:intpar}
\end{equation}
We finally express $\delta$ as
\begin{equation}
\label{eq:delta}
\frac{j_{s,{\rm NM}}(z_{\rm I})}
     {j_{s,{\rm NM'}}(z_{\rm I})} = \cosh{\delta} +
\frac{\rho_{\rm NM'} l_{\rm NM'}}{AR_{\rm I}} \delta \sinh{\delta}
\end{equation}
in terms of $j_{si}(z_{\rm I})$, the spin current at the interface $z_{\rm I}$ on the $i=$NM and NM$'$ sides as well as $\rho_{\rm NM'}$, $l_{\rm NM'} \equiv l_{\rm sf}^{\rm NM'}$ and $R_{\rm I}$. The relationship of the SML to the spin current discontinuity $j_{s,{\rm NM}}(z_{\rm I}) - j_{s,{\rm NM'}}(z_{\rm I})$ is nontrivial.

\begin{figure}[t]
\includegraphics[width=1.0\linewidth]{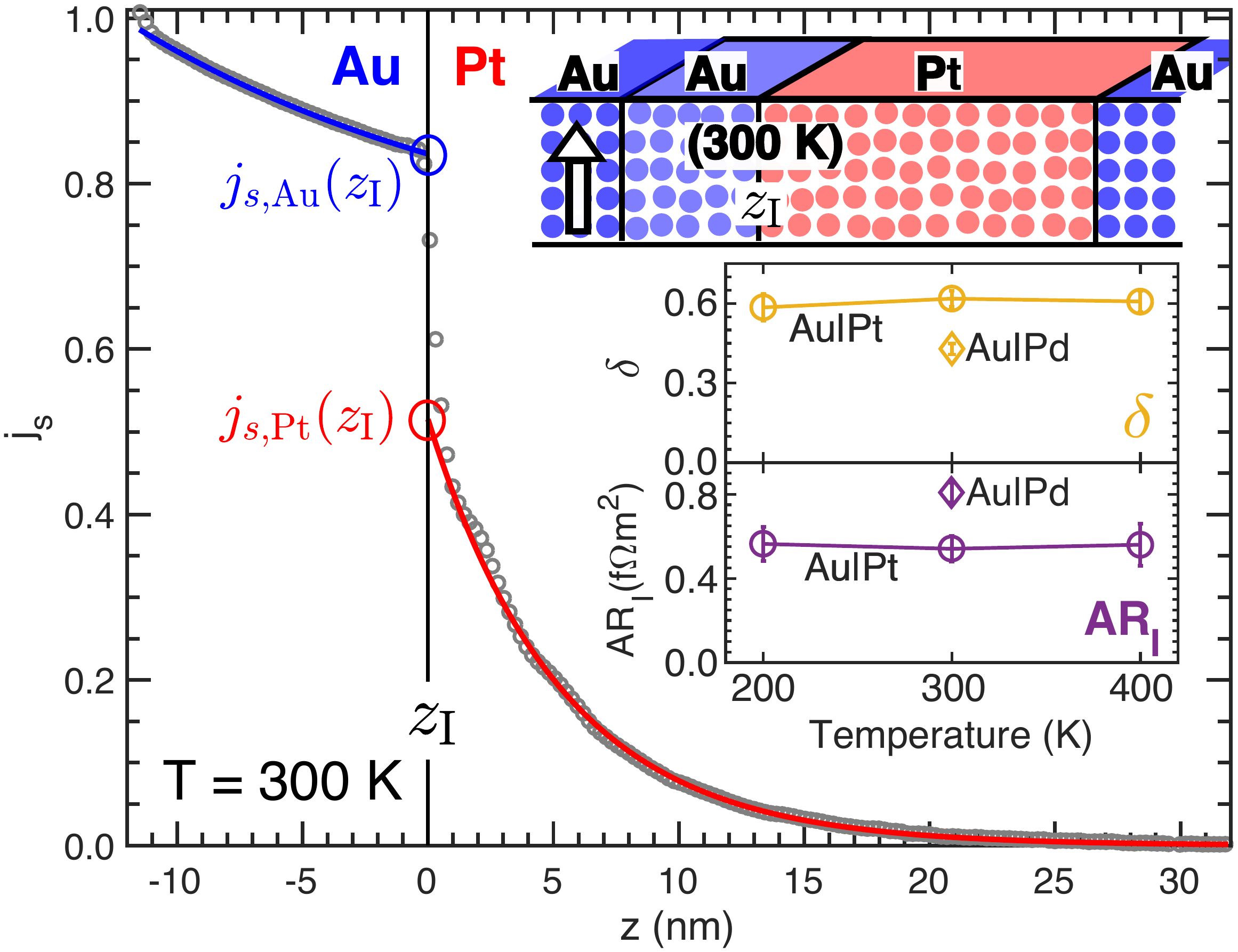}
\caption{A fully polarized spin-current $j_s$ injected at 300 K from the left Au lead into a Au(50)$|$Pt(140) bilayer sandwiched between Au leads decays exponentially in Au and in Pt; the numbers in brackets denote the number of atomic layers. The solid lines indicate fits for $j_s(z)$ in individual layers using solutions of the VF equations. Inset (top panel): schematic of the scattering region. (Middle and bottom panels): temperature dependence of the interface parameters $\delta$ (yellow) and $R_{\rm I}$ (purple) for a Au$|$Pt interface (circles). The corresponding parameters for Au$|$Pd at 300 K are included (diamonds).}
\label{fig1}
\end{figure}

{\color{red}\it {Au$|$Pt interface.---}}We illustrate our methodology in \cref{fig1} for a Au$|$Pt interface between ``room-temperature'' (RT) Au and Pt in which a Gaussian distribution of atomic displacements in a 7$\times$7 lateral supercell was used to reproduce the experimentally observed resistivities of each bulk material at T=300K, $\rho_{\rm Au}=2.3 \mu \Omega$cm and $\rho_{\rm Pt}=10.7 \mu \Omega$cm \cite{HCP90} for which $l_{\rm Au} \sim 80 \,$nm and $l_{\rm Pt} \sim 5.25 \pm0.05\,$nm \cite{Wesselink:prb19}. The empty grey circles in Fig.~\ref{fig1} represent $j_s(z)$ obtained \cite{footnote3} from the results of quantum mechanical scattering calculations \cite{Starikov:prb18} for a Au$|$Pt bilayer when a fully polarized spin current was injected into the bilayer from the left Au lead. The lattice constant of fcc Au is initially chosen to be that of Pt ($a'_{\rm Au}=a_{\rm Pt}=3.923$\AA) which does not affect the Au electronic structure qualitatively. The figure also shows the VF solutions in Au (blue curve) and Pt (red curve) found by fitting $j_s(z)$ far from the interface. The initial spatial decay of $j_s(z)$ is determined by $l_{\rm Au}$, the rapid decay in the vicinity of the interface is described in the semiclassical VF framework by the interface discontinuity and, after this abrupt decay, the spin current that survives in Pt decays to zero on a length scale described by $l_{\rm Pt}$. 

By fitting $j_s(z)$ to the solution of the VF equation, we obtain values of $j_{s,{\rm Au}}(z_{\rm I})$ and $j_{s,{\rm Pt}}(z_{\rm I})$. From the Landauer expression for the conductance in terms of the transmission matrices, $AR_{\rm I}=0.54 \pm0.03\,$f$\Omega$m$^2$ is directly determined leaving just $\delta$ as the only unknown parameter. Using a numerical root finder to solve \eqref{eq:delta}, we find $\delta=0.62\pm0.03$ where the error bar is evaluated from the uncertainties in the other parameters. 

The bulk parameters $\rho_{\rm Pt}$ and $1/l_{\rm Pt}$ are known to increase linearly with temperature \cite{HCP90, LiuY:prb15, Isasa:prb15a, WangL:prl16} but virtually nothing is known about the temperature dependence of interface parameters. We therefore calculate $AR_{\rm Au|Pt}$ and $\delta$ at 200 and 400~K and plot the results in Fig.~\ref{fig1} (inset). Within the error bars of the calculations, both parameters remain constant between 200 and 400 K. The temperature independence that we observe for $\delta$ is in agreement with the results of a very recent CPP-MR experimental study for Cu$|$Pt interface that shows $\delta$ to be nearly constant over the temperature range 0-300~K \cite{Freeman:prl18}.

\begin{figure}[t]
\includegraphics[width=1 \linewidth]{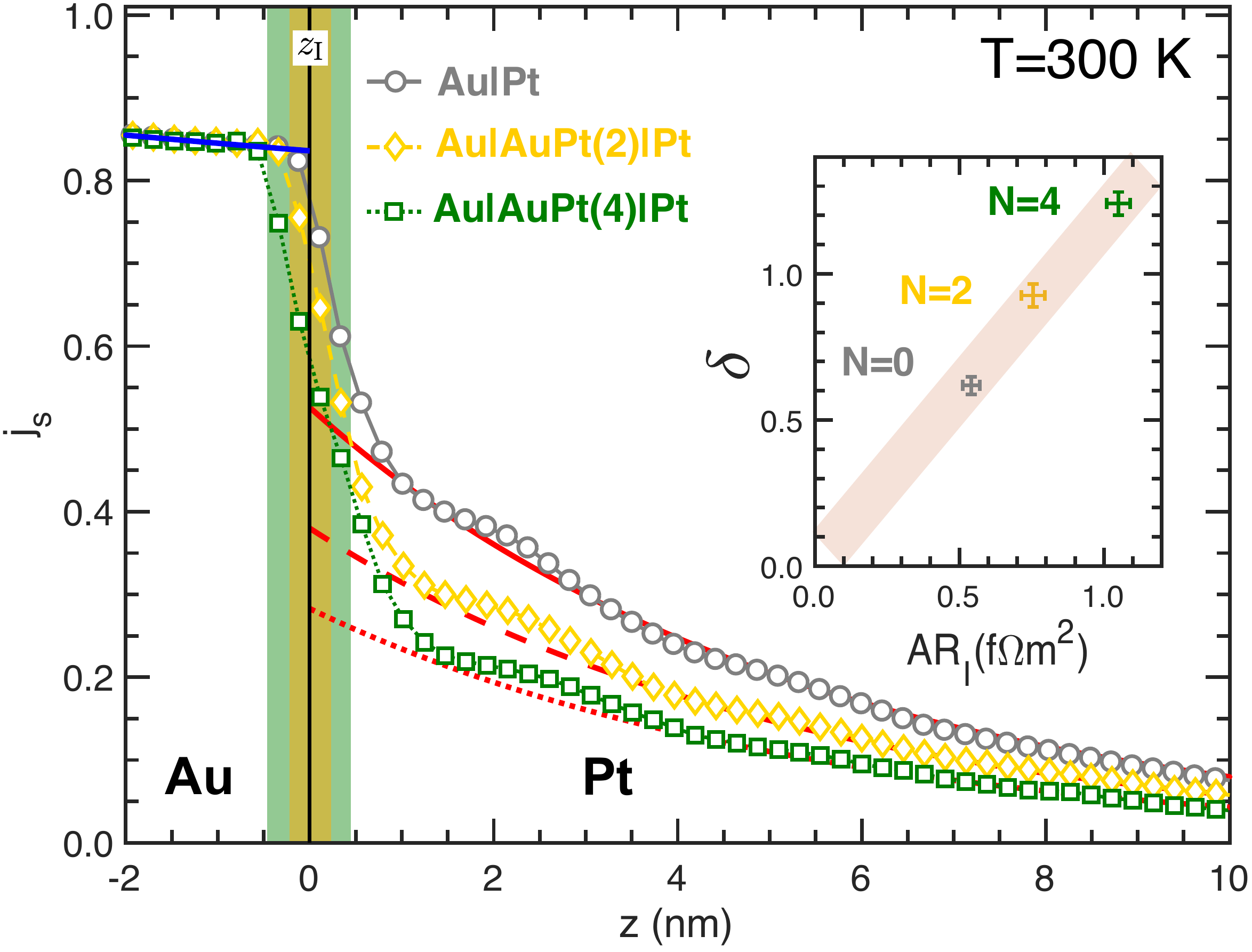}
\caption{A fully polarized spin current $j_s(z)$ is injected into a Au$|$Pt bilayer  with: a sharp interface (vertical black line), 2 layers of $\rm Au_{50}Pt_{50}$ interface (yellow shaded region) and 4 layers of $\rm Au_{50}Pt_{50}$ interface (green shaded region) between them. The calculated spin currents $j_s(z)$ for the three cases are shown as gray circles, yellow diamonds and green squares respectively. The solid blue line indicates a fit to the VF equation in Au. The solid, dashed and dotted red lines indicate fits to the VF equation in Pt for Au$|$Pt, Au$|\rm Au_{50}Pt_{50}(2)|$Pt and Au$|\rm Au_{50}Pt_{50}(4)|$Pt respectively. (Inset) $\delta$ vs $AR_{\rm I}$ for $N=0,2$ and 4 interface layers of mixed $\rm Au_{50}Pt_{50}$.
}
\label{fig2}
\end{figure}

{\color{red}\it {Interface Mixing.---}}Unlike the sharp interfaces between bulk Au and Pt we have considered so far, experimental interfaces are believed to comprise a few intermixed layers. To study the effect of interface mixing, we insert $N$ atomic layers of Au$_{50}$Pt$_{50}$ random alloy at the interface of the lattice-matched  Au$|$Pt bilayer \cite{footnote2}. The results for the spin current $j_s(z)$ and corresponding values of $\delta$ at 300 K are shown in \cref{fig2} for $N=0,2,4$. When the spin current from bulk Au approaches the mixed interface layers (yellow for $N=2$, green for $N=4$), then compared to the sharp interface, $j_s(z)$ decreases more and $\delta$ increases rapidly with increasing $N$ (inset). Electron scattering at a commensurable and clean Au$|$Pt interface only involves Bloch states with equal $\mathbf k_\|$ but intermixing (and thermal disorder) break momentum conservation and allow ${\bf k_\| \rightarrow k'_\|}$ scattering. The higher scattering rate results in a higher spin-flipping probability and hence a larger $\delta$ for the intermixed interfaces. Moreover, conduction electrons injected into Au are only weakly affected by SOC until they enter Pt where as $d$ states they become very susceptible to the large SOC. The interatomic mixing effectively increases the region where conduction electrons experience large SOC and therefore increases the SML. 
The interface resistance $AR_{\rm I}$ also increases monotonically as the disordered region increases in thickness suggesting that $\delta \propto AR_{\rm I}$ so $\rho_{\rm I}l_{\rm I} \sim$ const.

{\color{red}\it {Lattice Mismatch.---}}To study the effect of ${\bf k_\| \rightarrow k'_\|}$ scattering on its own, we re-examine the sharp (111) Au$|$Pt interface where both Au and Pt have their equilibrium bulk volumes, $a_{\rm Au}=4.078\,$\AA\ and $a_{\rm Pt}=3.923\,$\AA. A (111) oriented 5$\times$5 unit cell of Au matches to a (111) oriented $3\sqrt{3}\times3\sqrt{3}$ unit cell of Pt to better than 0.02\%. For this fully relaxed Au$|$Pt geometry, we repeat our calculations at 300 K and obtain $\delta=0.81\pm0.05$ compared to $\delta=0.62\pm0.03$ with commensurable Au. This calculation indicates that the ${\bf k_\| \rightarrow k'_\|}$ scattering does indeed lead to an increase of the interface SML. Our finding is in agreement with calculations for a Cu$|$Pd interface using Schep's ansatz \cite{Schep:prb97, footnote1} which indicated that $\delta$ increases on going from a sharp to a rough interface \cite{Belashchenko:prl16}.


{\color{red}\it {Au$|$Pd.---}}To examine the effect of changing the strength of the SOC, we apply the procedures described above to a commensurable Au$|$Pd interface, choosing $a'_{\rm Au}=a_{\rm Pd}=3.891$\AA. Corresponding to the experimental resistivity of Pd at 300 K, $\rho_{\rm Pd}=10.8~\mu\Omega\rm cm$ \cite{HCP90}, we find $l_{\rm Pd}=7.06\pm0.02$ nm and a value of $AR_{\rm Au|Pd}=0.81\pm0.05~{\rm f}\Omega\,{\rm m^2}$ which is much larger than the value $0.54\pm0.03~{\rm f}\Omega{\rm m^2}$ found for Au$|$Pt (inset \cref{fig1}, bottom panel). By substituting all the input parameters and their uncertainties into \eqref{eq:delta}, we extract a value of $\delta_{\rm Au|Pd}=0.43\pm0.02$ (\cref{fig1}, inset). Compared to Au$|$Pd, the larger SOC in Pt leads to a larger value of $\delta$ for Au$|$Pt. 
Our results for $AR_{\rm I}$ and $\delta$ for Au$|$Pd interfaces are in good agreement with theoretical estimates made by Belashchenko {\it et al.} combining Schep's ansatz \cite{Schep:prb97} with calculations for ballistic Cu$|$Pd interfaces  \cite{Belashchenko:prl16}. Cu and Au have very similar electronic structures and the very different SOC of the filled 3$d$ and 5$d$ states below the Fermi level is not expected to play a major role.

\begin{figure}[!t]
\includegraphics[width=1.0\linewidth]{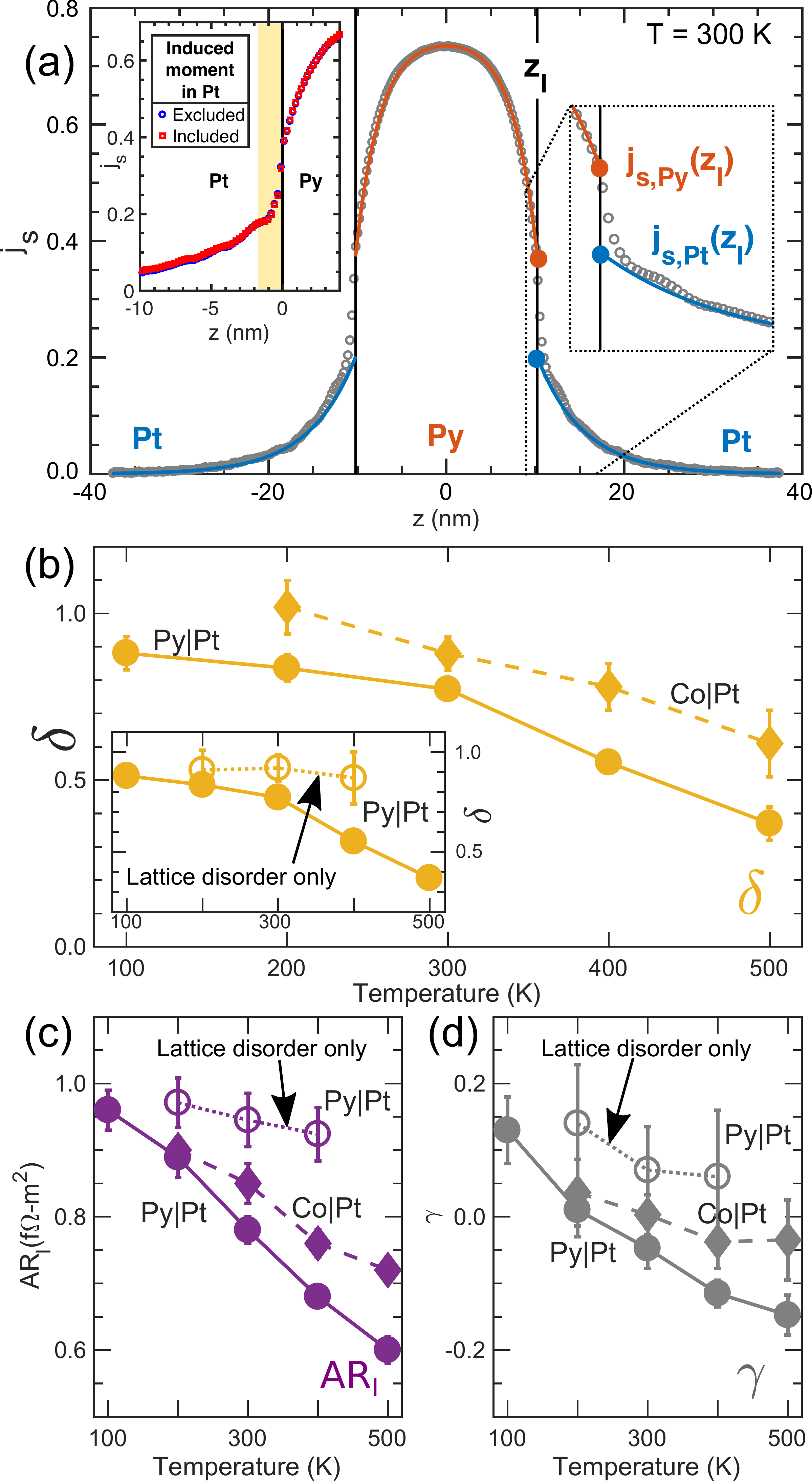}
\caption{(a) Open circles: spin current $j_s(z)$ through a Pt$|$Py$|$Pt trilayer calculated for $T=300\,$K. The solid blue (orange) curve is a fit to the VF equations in bulk Pt (Py). These fits are extrapolated to the interface $z_{\rm I}$ to obtain the values $j_{s,\rm Pt}(z_{\rm I})$ and $j_{s,\rm Py}(z_{\rm I})$ shown in detail in the right inset. The left inset shows the spin current with (red) and without (blue) proximity induced moments in Pt. (b) $\delta$ for Py$|$Pt (circles, solid lines) and Co$|$Pt (diamonds, dashed lines) plotted as a function of temperature. (Inset) $\delta$ for Py$|$Pt compared with results with only lattice disorder in Py$|$Pt (open circles, dotted lines). (c) and (d) Interface parameters $AR_{\rm I}$ and $\gamma$ for Py$|$Pt (circles, solid and dotted lines) and Co$|$Pt (diamonds, dashed lines) plotted as a function of temperature. The dotted lines show the results for Py$|$Pt with only lattice disorder.
}
\label{fig3}
\end{figure}

{\color{red}\it {FM$|$Pt interfaces.---}}We developed an analogous procedure to study FM$|$NM interfaces. Compared to the NM$|$NM$'$ case, two additional parameters enter: spin asymmetry parameters $\beta=(\rho_{\downarrow}-\rho_{\uparrow})/(\rho_{\downarrow}+\rho_{\uparrow})$ for the bulk FM and $\gamma=(R_{\downarrow}-R_{\uparrow})/(R_{\downarrow}+R_{\uparrow})$ for the interface. To avoid interfaces between a lead and Py or Co, we considered symmetric NM$|$FM$|$NM scattering geometries and studied them by passing an unpolarized charge current through them. The appropriate boundary conditions are that both the spin accumulation and spin current vanish at $z=\pm \infty$ and the analysis results in two implicit equations containing the discontinuity in the spin current at the FM$|$NM interface as described by $j_{s,{\rm FM}}(z_{\rm I})$ and $j_{s,{\rm Pt}}(z_{\rm I})$ as well as the eight transport parameters 
$\rho_{\rm NM}$, 
$l_{\rm sf}^{\rm NM}$,
$\rho_{\rm FM}$, 
$l_{\rm sf}^{\rm FM}$, 
$\beta_{\rm FM}$,
$R_{\rm I}$,
$\delta$ and $\gamma$.  \cref{fig3}(a) illustrates the spin current $j_s(z)$ that we calculate for a Pt$|$Py$|$Pt trilayer at 300 K. The five bulk parameters are determined independently as well as $AR_{\rm I}$ obtained from the Landauer formula leaving us with two equations and two unknowns, $\delta$ and $\gamma$, to be determined. 

 This procedure was applied to Py$|$Pt and Co$|$Pt interfaces assuming completely relaxed geometries and 8$\times$8 interface unit cells of (111) Py or Co matched to 2$\sqrt{13} \times 2 \sqrt{13}$ interface unit cells of Pt. Thermal lattice and spin disorder were taken into account as described in Refs.~\onlinecite{LiuY:prb11, LiuY:prb15, Starikov:prb18, Wesselink:prb19}. $\rho_{\rm Pt}$ and $l_{\rm Pt}$ were already determined above and the appropriate corresponding calculations were performed for bulk Py and Co \cite{LiuY:prb15, Gupta:19}.

The temperature dependence of the three interface parameters that we extract for Py$|$Pt and Co$|$Pt interfaces is summarized in \cref{fig3}(b-d). $AR_{\rm I}$ and $\gamma$ are seen to decrease monotonically with temperature for both interfaces. $\gamma$ is found to vary in a small range between $\pm 0.15$ for Py$|$Pt and between $\pm 0.03$ for Co$|$Pt. $\delta$, the main focus of our interest, decreases monotonically and substantially with temperature for both interfaces. $\delta_{\rm Co|Pt}$ is larger than $\delta_{\rm Py|Pt}$ for all temperatures in the range 200-500 K. This temperature dependence contrasts starkly with the temperature independence we found for Au$|$Pt. We speculate that it is the variation of the spin disorder associated with the FM magnetization that affects the interface parameters most. Consistent with this is our finding that $\delta$ and $AR_{\rm I}$ are larger for Co$|$Pt than for Py$|$Pt. With a higher Curie temperature Co is more ordered at any given temperature than Py. 

To test this hypothesis, we repeated the $T=200, 300$ and $400\,$K calculations for Py$|$Pt keeping the atomic spins ordered and including only lattice disorder in Py. The results for the three interface parameters with only lattice disorder are included in \cref{fig3} (open circles, dotted lines) for comparison. With only lattice disorder included, we find that the Py$|$Pt interface parameters decrease much more slowly with temperature. This weak variation can be attributed to the lattice disorder, but the decrease is much smaller compared to that brought about by spin disorder. In the low-temperature limit, we also expect $\delta$ to be smaller for Py$|$Pt because this interface is less abrupt than Co$|$Pt owing to Py's intrinsic disorder. SOC-induced interface splittings are smeared out by alloy disorder in Py compared to Co, leading to smaller $\delta$.
Lastly, we found that proximity induced magnetization of Pt by Co or Py has no effect on the interface parameters within the error bars of the calculations; see the left inset to \cref{fig3}(a).

{\color{red}\it {Summary.---}}
First-principles scattering theory and a recently developed local current scheme have been used to study how spin currents propagate through interfaces between two nonmagnetic (Au$|$Pt and Au$|$Pd) materials and between a ferromagnetic and a nonmagnetic (Py$|$Pt and Co$|$Pt) material at finite temperatures. By extracting values of $\delta$, $R_{\rm I}$ and $\gamma$ we could study how $\delta$ depends on various properties of the interfaces and temperature. For nonmagnetic interfaces, we found that $\delta$ and $AR_{\rm I}$ remain unchanged over a wide range of temperature and found values of $AR_{\rm I}$ that are in remarkably good agreement \cite{footnote1} with an {\it ansatz} introduced more than twenty years ago by Schep {\it et al} \cite{Schep:prb97}.  

$\delta_{\rm Au|Pt}$ was found to be larger than $\delta_{\rm Au|Pd}$ owing to the larger SOC in Pt, indicating a direct link between the magnitude of $\delta$ and SOC strength of NM metals. An incommensurable Au$|$Pt interface with relaxed Au and Pt lattices has a substantially larger $\delta$ than the lattice matched interface. Mixing at an interface also leads to larger values of $\delta$. Thus to minimize $\delta$, lattice matched and clean interfaces should be targeted in experiments to avoid momentum-nonconserving scattering of conduction electrons.

FM$|$Pt interface parameters decrease strongly with increasing temperature. This dependence stems directly from the magnetization of the FM. Co is a stronger FM than Py and we find that $AR_{\rm Co|Pt}$ and $\delta_{\rm Co|Pt}$ are larger than $AR_{\rm Py|Pt}$ and $\delta_{\rm Py|Pt}$ for all temperatures. By turning off spin disorder in Py$|$Pt, the variation of interface parameters with temperature becomes negligible.

{\color{red}\it {Acknowledgements.---}}This work was financially supported by the ``Nederlandse Organisatie voor Wetenschappelijk Onderzoek'' (NWO) through the research programme of the former ``Stichting voor Fundamenteel Onderzoek der Materie,'' (NWO-I, formerly FOM) and through the use of supercomputer facilities of NWO ``Exacte Wetenschappen'' (Physical Sciences). K.G. acknowledges funding from the Shell-NWO/FOM “Computational Sciences for Energy Research” PhD program (CSER-PhD; nr.~i32; project number 13CSER059) and is grateful to Yi Liu for help in starting this work and to S. Wildeman for helpful discussions. The work was also supported by the Royal Netherlands Academy of Arts and Sciences (KNAW). Work in Beijing was supported by the National Natural Science Foundation of China (Grant No. 61774018), the Recruitment Program of Global Youth Experts, and the Fundamental Research Funds for the Central Universities (Grant No. 2018EYT03).

%


\begin{thebibliography}{58}
\bibitem [{foo({\natexlab{a}})}]{footnote5}%
  \BibitemOpen
  \bibinfo {note} {The results we will describe are valid in the linear response regime that should be accessible using nonlocal spin-injection experiments. When properties are determined in switching experiments i.e., from spin pumping and ISHE measurements or from SHE and SOT measurements care must be taken to ensure   that nonlinear effects can be neglected \cite{Demidov:natc17, Borisenko:apl18} \mbox{} }\BibitemShut {NoStop}%
\bibitem [{foo({\natexlab{b}})}]{footnote3}%
  \BibitemOpen
  \bibinfo {note} {{The spin current flowing in the $z$ direction and polarized along the global quantization axis of the system is defined in terms of the nonconserved spin as $j_s(z) \equiv\hat z\cdot\langle\sigma\otimes\mathbf v+\mathbf v\otimes\sigma\rangle/2$. It is normalized w.r.t. the charge current density  and is averaged over the $xy$ plane perpendicular to the transport direction. For more details, see Ref.~[\onlinecite{Wesselink:prb19}].}}\BibitemShut {Stop}%
\end{thebibliography}
\end{document}